\PassOptionsToPackage{twoside=false}{geometry}
\documentclass[acmlarge,screen,nonacm]{acmart}

\usepackage{color}  

\makeatletter
\patchcmd{\@mkauthors@i}% <cmd>
  {\@typeset@author@line\par}% <search>
  {\@typeset@author@line\vspace{0.05in}\par\noindent{\normalsize \@date}\par}% <replace>
  {}{}% <success><failure>
\makeatother

\setcopyright{none}
\copyrightyear{2025}
\acmYear{2025}

\begin{document}

\title{Extending MovieLens-32M to Provide New Evaluation Objectives}

\author{Mark D.\ Smucker}
\affiliation{%
  \institution{University of Waterloo}
  \city{Waterloo}
  \state{Ontario}
  \country{Canada}}
\email{mark.smucker@uwaterloo.ca}
\orcid{0000-0003-4968-6405}

\author{Houmaan Chamani}
\affiliation{%
  \institution{University of Waterloo}
  \city{Waterloo}
  \state{Ontario}
  \country{Canada}}
\email{houmaan.chamani@uwaterloo.ca}
\orcid{0009-0009-9725-4309}

\date{April 26, 2025\\
This is a preprint. The definitive version was published
in SIGIR 2025, \url{https://doi.org/10.1145/3726302.3730328}.}

\begin{abstract}

Offline evaluation of recommender systems has traditionally treated
the problem as a machine learning problem.  In the classic case of
recommending movies, where the user has provided explicit ratings of
which movies they like and don't like, each user's ratings are split
into test and train sets, and the evaluation task becomes to predict
the held out test data using the training data.  This machine learning
style of evaluation makes the objective to recommend the movies that a
user has watched and rated highly, which is not the same task as
helping the user find movies that they would enjoy if they watched
them. This mismatch in objective between evaluation and task is a
compromise to avoid the cost of asking a user to evaluate
recommendations by watching each movie.  As a resource available for
download, we offer an extension to the MovieLens-32M dataset that
provides for new evaluation objectives. Our primary objective is to
predict the movies that a user would be interested in watching,
i.e.\ predict their watchlist.  To construct this extension, we
recruited MovieLens users, collected their profiles, made
recommendations with a diverse set of algorithms, pooled the
recommendations, and had the users assess the pools.
This paper demonstrates the feasibility of using pooling to construct
a test collection for recommender systems.  Notably, we found that the
traditional machine learning style of evaluation ranks the Popular
algorithm, which recommends movies based on total number of ratings in
the system, in the middle of the twenty-two recommendation runs we
used to build the pools.  In contrast, when we rank the runs by users'
interest in watching movies, we find that recommending popular movies
as a recommendation algorithm becomes one of the worst performing
runs.
It appears that by asking users to assess their personal
recommendations, we can alleviate the issue of popularity bias in the
evaluation of top-n recommendation.

\end{abstract}

\begin{CCSXML}
<ccs2012>
   <concept>
       <concept_id>10002951.10003317.10003359</concept_id>
       <concept_desc>Information systems~Evaluation of retrieval results</concept_desc>
       <concept_significance>500</concept_significance>
       </concept>
   <concept>
       <concept_id>10002951.10003317.10003359.10003360</concept_id>
       <concept_desc>Information systems~Test collections</concept_desc>
       <concept_significance>500</concept_significance>
       </concept>
   <concept>
       <concept_id>10002951.10003317.10003347.10003350</concept_id>
       <concept_desc>Information systems~Recommender systems</concept_desc>
       <concept_significance>500</concept_significance>
       </concept>
 </ccs2012>
\end{CCSXML}

\ccsdesc[500]{Information systems~Evaluation of retrieval results}
\ccsdesc[500]{Information systems~Test collections}
\ccsdesc[500]{Information systems~Recommender systems}

\keywords{Recommender Systems; Test Collection; Pooling}

\maketitle

\section{Introduction}

Recommendation systems have long been evaluated by collecting a large
number of individuals' ratings for items, and then dividing these
ratings into train and test sets to see how effective a recommendation
algorithm is.  Early work focused on prediction of test set ratings.
More recent work has focused on the ranking performance of algorithms
used for top-n recommendation with the test set ratings functioning as
relevance judgments.

A complaint about this approach to recommendation system test
collection construction is that the collections are typically created
as the by-product of a running recommendation system.  For example,
the MovieLens datasets~\cite{movielens} are the movie ratings of its
users over a period of more than two decades.  As noted
by~\citet{grouplens94}, using these sparse ratings matrices in this
manner could lead to a bias with the items being rated coming
predominately from the recommendation algorithms used by the system.
Likewise, \citet{konstan2012:overview} explain that by using a user's
existing ratings, we are measuring a recommendation algorithm's
ability to recommend items already known to the user, whereas the goal
of a recommender is presumably to find items unknown to the user.

\citet{bellogin17:biases} highlight two problems with using
information retrieval effectiveness measures in conjunction with a
train/test split evaluation approach: sparsity and popularity bias.
When we use test set ratings as relevance judgments, the judgments are
likely to be incomplete (sparse), especially for profiles with smaller
numbers of ratings.  \citet{buckley04:incomplete} showed that the more
incomplete relevance judgments are, the less likely that we can
correctly order the effectiveness of ranking algorithms.
The other problem that results from using existing ratings
as relevance judgments for evaluation of top-n recommendation
is \textit{popularity bias}.  Popular items by definition are those
items that users are more likely to have judged. When we use the
user's existing profile to select test items, we are more likely to
pick popular items, and thus recommending popular items
performs better than seems reasonable.

Finally, \citet{rossetti16:online} have shown that offline
evaluation of recommendation systems using a train/test split
methodology may not agree with actual user preferences when compared
to an evaluation that asks the user to assess the recommended items.

To address these issues, we used traditional information retrieval
(IR) test collection construction techniques to create an extension to the
MovieLens-32M (ML-32M) dataset.  After receiving clearance from our
university's research ethics board, we recruited movielens.org
users to be participants in a research study to investigate the
feasibility of using pooling for creating the relevance judgments for
a recommender systems test collection.

Pooling has commonly been rejected for recommendation test collections
as being too expensive or infeasible.  For example, in the case of
movies, if our objective function is to predict movies that the user
will enjoy watching, then to obtain our relevance assessments, it will
require us to obtain all of the unseen recommended movies and then
have the user watch hundreds of hours of movies.

Our objective function is to predict unrated movies that the user is
interested in watching.  The usage case for this objective is a user
that wants to find movies to add to a ``watchlist'' or queue of
movies, which are common features of online streaming services.  For
our objective function, our participants assessed relevance given a
poster image, title, year, plot summary, and other useful information
such as actors and director.  While some participants said that they
would normally watch a movie trailer to help them decide on their
interest, we instructed them to make their judgments based on the
information we provided.  Our participants rated movies at an average
rate of one every 20.3 seconds.

In addition to our primary objective function, for which we asked our
participants to tell us their interest level in watching a movie, our
collected relevance assessments included familiarity and predicted
ratings for unseen movies and ratings for already seen movies (users'
ratings profiles do not include all seen movies and thus seen movies
can still be recommended).  With this additional information, we can
produce many different sets and types of relevance judgments,
i.e.\ many different objectives.  For example, we can restrict the
relevance judgments to unfamiliar movies to see
which algorithms can find such movies.

Our extension to ML-32M consists of 51 participant ratings profiles
and 31,236 relevance judgments for movies that our participants had
not previously rated and reported to movielens.org.  We used our
extension to evaluate the performance of the diverse set of algorithms
used to construct the judgments pools, and when compared to a
traditional approach that does a random 80/20 split of the ratings
profiles as train/test sets, we show:

\begin{itemize}

\item Pooling is a feasible approach to building a test collection for
  recommender systems.
  
\item Using pooling and interest-in-watching preferences with the
  \textit{compatibility} effectiveness measure to rank top-n
  recommendation runs, pushes the Popular run, which recommends based
  on number of ratings, near the bottom of the ranking, while a
  traditional train/test split finds Popular to be better than half of
  the other runs we used.
  
\item Our participants are different from a random ML-32M user and
  this contributes to a difference in evaluation results, which is
  likely a positive effect given that our participants appear to be
  serious users of movielens.org.

\end{itemize}

Based on our findings, we recommend using our interest-in-watching
preferences for offline evaluation with ML-32M for its apparent
ability to reduce or remove popularity bias from the evaluation.  We
detail our recommendations in Section~\ref{sec.recs.for.use}.

Our extension to MovieLens-32M is available for researchers at
{\color{blue}\url{https://uwaterlooir.github.io/datasets/ml-32m-extension}}.

\section{Related Work}

\citet{mclaughlin04:ux} identified that the use of prediction accuracy
rather than IR measures for ranking evaluation was flawed and wrote
that a key issue that had allowed the field to make this mistake was
using offline evaluation that had not been validated to align with
user experience.  In other words, if researchers had shown the
recommendations to the users, the users could have easily told them
that they were filled with many significant mistakes. We follow their
advice by directly asking our participants to assess the quality of
recommended movies.

\citet{castells22:offline} provide an excellent comparison of IR
offline evaluation methodology and the traditional machine learning
evaluation approach commonly used for recommender systems.  

\citet{klimashevskaia24:popbias} comprehensively survey the issue of
popularity bias in recommender systems.  Popularity bias exists both
in the recommendations made by algorithms and in the application of
information retrieval effectiveness measures to top-n recommendations.
In this paper, our interest in popularity bias is limited to the bias
of effectiveness measures, but popularity bias in the recommendation
algorithms that we used to construct our judging pools could harm the
reusability of our test collection extension.  In addition to
identifying the issues of sparsity and popularity
bias, \citet{bellogin17:biases} provide methods to ameliorate these
issues when a traditional train/test split method is used for
evaluation of recommender systems with information retrieval
effectiveness measures.  We believe that the method of pooling and
asking the users themselves to assess the recommendations is a more
direct solution to the problem of popularity bias in evaluation.
Whatever the user assesses as preferred is what is preferred
regardless of whether it is a popular movie or not.  As for the issue
of sparsity, i.e.\ the incompleteness of relevance judgments, it
remains a potential issue with pooling and one that we leave to future
work.

\citet{canamares18:follow} show that popularity bias in evaluation
can be understood in terms of how users discover, consume, and rate
items.  Their analysis shows that the average rating of an item
should be a better predictor of user preference that popularity.  Our
results support their analysis.  As we show in
Figure~\ref{fig.cran.v.trad}, pooling-based evaluation ranks the Bias
algorithm of LensKit before the Popular algorithm, while a traditional
train/test split evaluation places Popular as significantly better
than Bias.  When used for ranking, the Bias algorithm produces
recommendations based on average rating with a correction (damping)
for items that have few ratings.  \citeauthor{canamares18:follow} tested
their analytical results by creating an unbiased, crowdsourced
dataset.  Their dataset differs from our work in that they randomly
selected songs and randomly assigned songs to people to judge while we
retain the standard use of collected user ratings for making
predictions.  Our study participants judged a diverse set of
recommended movies that aims to avoid the bias issues inherent in the
traditional train/test approach to evaluation.  Pooling should also
help to reduce sparsity of judgments, which we would not expect to be
possible with judging random items. 

\citet{abdollahpouri21:upd} note that different users have different
preferences, and in particular, some users want to be recommended
popular blockbusters, while at the other extreme, some users have
disdain for the popular and prefer niche
movies. As such, \citeauthor{abdollahpouri21:upd} propose User
Popularity Deviation (UPD) as a measure of how well recommendations
match the distribution of popularity that a user has in their existing
ratings profile. While such a measure could still be used with
pooling-based evaluation, if the pools are diverse enough, we should
be able to directly trust the preferences we collect from users when
they judge the pools.

\citet{sun23:fresh} explains the importance of making train/test
splits such that the training data user-item interactions occur in time
before the interactions used for evaluation.  Our approach respects
this important split.

Our work is most related to that of \citet{rossetti16:online}, who
conducted a 100 user study to evaluate the degree to which traditional
offline evaluation (all-but-one) agreed with users' assessments of the
pooled recommendations of four different algorithms at a depth of 5.
\citet{rossetti16:online} found that conclusions regarding which
algorithms were better, changed between their traditional offline
evaluation and the pooling-based evaluation.

In many regards, their study was similar to ours, but also different.
They also used summaries of the key movie information in place of
having people watch the actual movies, asked their users about
familiarity and whether the movie had already been seen, and asked users
about their interest in watching the movie.  Unlike us, they did not collect
MovieLens scaled ratings, and they did not recruit movielens.org users
and instead had students browse a collection of movies and rate them.

In contrast to \citet{rossetti16:online}, our goal was to create an
extension to ML-32M to allow for improved offline evaluation.  As
such, we utilized a much larger set of runs to generate the judgment
pools and had our participants judge much deeper into the pools.
Likewise, they utilized MovieLens-1M, which is out of date, while we
used ML-32M, which had been created immediately preceding the
recruitment of our participants.  In effect, MovieLens-32M was frozen
and then our participants received judgment pools that were no more
than 2 months out of date, and as such captures as best as possible
the opinions of our participants at the same point in time as data in
ML-32M.

While our assessors are primary assessors, i.e.\ the people with the
information need, \citet{lu21:secondary} investigated the use of
secondary assessors to perform assessments and found evidence that
it is feasible for movie recommendations.

\section{Creating the Test Collection}

In this section we explain how we created our new test collection for
offline evaluation of recommendation algorithms.  Space limits us to
presenting the key aspects of the test collection's construction.
Full details are provided by \citet{chamani24:thesis}.

\subsection{Summary of Design}

We transformed and extended
MovieLens-32M\footnote{\url{https://grouplens.org/datasets/movielens/32m/}}
(ML-32M) to create our new test collection.  After obtaining ethics
clearance from the University of Waterloo's Research Ethics Board, we
recruited participants from movielens.org.  Participants provided us a
download of their movielens.org ratings.  We appended their ratings to
a transformed version of ML-32M and then generated recommendations for
each participant using a diverse set of algorithms.  For each
participant, we pooled their recommendations and provided a website
where the participant could assess each recommendation.

Assessment of movies was split into two phases. Phase 1 involved the
judging of an average of 152.7 movies, and for those that completed
phase 2, each participant judged an average of 670.5 movies in total.
Phase 1 had a minimum pool depth of 10, and phase 2 reached a minimum
pool depth of 50.  Actual depth of pooling varied by participant to
enable us to collect similar amounts of judgments from each
participant.  We only provide data for the 51 participants who
finished phase 2.

Participants assessed each movie on their familiarity, desire to watch
the movie, and for seen movies, their rating of the movie, and for
unseen movies, their predicted rating.  In addition to the
recommendations, we put a random sample of their provided ratings into
the judgment pool to allow us to verify and test the quality of their
provided judgments.  With the collected judgments, we produced a
variety of different sets of relevance judgments to allow for many
different evaluation objectives.

We provide further details in the remainder of this section.

\subsection{Participants and Remuneration}
\label{sec.participants}

With the assistance of the GroupLens research group, we recruited
participants via the movielens.org website from Oct 17, 2023 to Dec 6,
2023 via the placement of a text banner inviting participation.
Interested people then clicked to a screening questionnaire.  We
required participants to be 18 years or older and to be residents of
Canada or the USA to enable us to remunerate them for their time.

Of the 360 people who signed up for the study, 271 passed screening,
130 gave consent, 113 provided their ratings profiles from
movielens.org, 107 submitted demographics, 103 passed a quiz about the
assessing instructions, 97 completed phase 1, 77 asked to do phase 2,
57 finished phase 2, and after removing participants that did not appear
to have an existing profile in ML-32M, we had 51 participants.  We
removed participants without an apparent existing profile in ML-32M to
eliminate people who may have joined movielens.org with the goal of
participating in the study to obtain its remuneration.

We estimated that phase 1 of the study would take participants
approximately 2 hours.  We remunerated these participants CAD\$40 or
USD\$30.  For phase 2, we tracked the time spent making assessments
and remunerated these participants CAD\$20/hour or USD\$15/hour with
their time spent rounded up to the nearest hour.  For both phases,
remuneration was in the form of Amazon.ca/Amazon.com e-gift cards.  We
also made pro-rated payments to participants who only partially
completed the phases.  In total, we spent CAD\$9,079.54.

For later analyses in the paper (Section~\ref{sec.comparison}), we
select 10K existing ML-32M users for analysis, which we will refer to
as ML-32M-10K.  These users were selected randomly from users that had
at least 20 ratings of 4.0 or greater in our final dataset.
\citet{chamani24:thesis} provides a detailed analysis of our study
participants compared to these 10K ML-32M users.  Notable differences
to other MovieLens users include the following findings.

Compared to the MovieLens-1M dataset, which included demographics, our
participants are a bit older (average age 36.9 years vs.\ 30.6 years)
and more identify as men (80.4\% vs.\ 71.7\%).  While ML-1M includes
people less than 18 years old, we only have people 18 years or older.

Our participants have rated many more movies.  The average ML-32M-10K
user has rated 190.9 movies, while our 51 participants have rated on
average 1425.1 movies.  Like ML-32M-10K, the distribution of movies
rated is skewed.  The median number of ratings for ML-32M-10K is 102
compared to 849 for our participants.

When rating movies, our participants have a more centered ratings
distribution with an average rating of 3.2 compared to 3.7 for
ML-32M-10K. In addition, our participants are less likely to rate a
movie 5 stars out of 5 with only 8.7\% of ratings being 5 stars
compared to 14.4\% for the ML-32M-10K users.

Our participants are actively rating movies while most MovieLens users
are seen in a given year and not again.  While we did not have the
timestamps of our participants' ratings, \citet{chamani24:thesis}
mapped our participants to their likely profile in ML-32M and
estimated that on average our participants had been using MovieLens
for 7.8 years in comparison to 10.5 months for ML-32M-10K users, and
estimated the median years of use of our participants at 5.2 years
compared to 15.8 hours for the ML-32M-10K users.

\subsection{MovieLens 32M Transformation}
\label{sec.data}

As it took an extended time to recruit participants, we processed them
in three batches.  For each batch, we appended their ratings profiles
to the existing ML-32M dataset and then transformed the data before
producing recommendations for the participants to judge.  We are
releasing the final transformed version of ML-32M to which our final
51 participants' data is appended.

As we needed to be able to show each participant information
concerning the movies, we
joined the ML-32M dataset with  
TMDB\footnote{\url{https://www.themoviedb.org/}} using the TMDB
API. If TMDB identified the movie as ``adult'' or as a TV show/series,
we excluded the item.  In addition, we found that some of the movies
in ML-32M had an invalid TMDB id, and if we could not manually find
the movie in TMDB, we excluded it.

We then applied 10-core filtering to the movies, i.e.\ we removed all
movies with fewer than 10 ratings.  We then applied 20-core filtering
to the users and only kept users with 20 or more ratings.  After the
20-core filtering, we applied another 10-core movie
filtering.

Given that we had recruited our participants from movielens.org, it
was reasonable to expect that most of them would have had an existing
profile in ML-32M, for ML-32M contains ratings up through Oct 12,
2023, and we began recruiting participants on Oct 17, 2023.  Given the
nature of many recommendation algorithms, having a duplicate of a
participant's profile in the dataset would affect their behavior as
the duplicate would be found as the most similar profile. To avoid
this issue, we utilized LensKit's UserUser-knn algorithm and used it
to compute the similarity between our participants and the existing
profiles.

For each participant, we found the profile with the highest similarity
and considered it a candidate match for the participant. Our
code output all profile matches with similarity greater
than or equal to 0.9 or the sole profile with a maximum similarity
below 0.9. We did not find any participant to have multiple high
similarity matches, i.e. over 0.9 similarity with multiple profiles.
We manually reviewed profile matches from lowest similarity up through
the matches in the 0.90-0.93 range. The matched profiles with
similarity of 0.85 or greater were clearly the same person in all
cases. In some cases we did consider profiles with lower scoring
matches to also be the same person, but these were rare. For all three
batches of participants, we confirmed that 103 profiles belonged to
our participants, and we removed these matching profiles from ML-32M.  

To exclude participants who may have joined movielens.org merely to
participate in our study, and thus with a possible goal
of collecting remuneration from us, we eliminated from the final
dataset all participants lacking a matching ML-32M profile with a
similarity greater than 0.85.

ML-32M contains 32,000,204 movie ratings from 200,948 users over
87,585 movies.  After our transformation and with our final 51
participants, we have 31,741,309 explicit ratings from 200,727 users
over 31,272 movies.

In addition to creating an explicit ratings dataset, we also produced
what we term an \textit{implicit} dataset.  The implicit dataset is
formed by taking the explicit dataset and only keeping ratings greater
than or equal to 4, and then by removing any users with fewer than 5
ratings.  This is the same transformation used by
\citet{liang18:multivae} (MultiVAE), \citet{steck19:ease} (EASE), and
\citet{wu16:cdae} (CDAE), which are some of the algorithms we utilize
to generate the recommendation pools.  The implicit dataset has
15,840,681 ratings from 198,762 users over 30,545 movies.

\subsection{Generating Diverse Recommendations}

We used LensKit~\cite{ekstrand20:lkpy} and
RecBole~\cite{recbole[1.0],recbole[2.0],recbole[1.2.0]} to generate
recommendations for our participants.  Our goal was to produce a wide
variety of recommendations using different algorithms including
non-personalized baselines, user and item knn based algorithms, matrix
factorization approaches, and modern neural network and related
algorithms. Table~\ref{tab.algs} shows the 22 algorithms that we used.

\begin{table*}
    \centering
    \caption{The twenty-two algorithms used to generate recommendation pools.}
    \label{tab.algs}
    \begin{tabular}{llll p{2.6in}}
      \toprule
      Run Name & Package & Algorithm & Dataset & Parameters (non-default) \\ \hline
      \midrule
      Popular & LensKit & basic.Popular & explicit & \textasciitilde \\ 
        Bias & LensKit & bias.Bias & explicit & damping=5 \\ 
        IIEx\_30\_2\_001 & LensKit & item\_knn.ItemItem & explicit & nnbrs=30, min\_nbrs=2, min\_sim=0.01 \\ 
        IIEx\_30\_10\_005 & LensKit & item\_knn.ItemItem & explicit & nnbrs=30, min\_nbrs=10, min\_sim=0.05 \\ 
        IIEx\_30\_30\_005 & LensKit & item\_knn.ItemItem & explicit & nnbrs=30, min\_nbrs=30, min\_sim=0.05 \\ 
        IIIm\_1\_1\_0001 & LensKit & item\_knn.ItemItem & implicit & nnbrs=1, min\_nbrs=1, min\_sim=0.001,\newline feedback=implicit \\ 
        IIIm\_120\_15\_0001 & LensKit & item\_knn.ItemItem & implicit & nnbrs=120, min\_nbrs=15, min\_sim=0.001,\newline feedback=implicit \\ 
        UUIm\_30\_2\_001 & LensKit & user\_knn.UserUser & implicit & nnbrs=30, min\_nbrs=2,\newline min\_sim=0.01,feedback=implicit \\ 
        UUEx\_30\_2\_01 & LensKit & user\_knn.UserUser & explicit & nnbrs=30, min\_nbrs=2, min\_sim=0.1 \\ 
        UUEx\_30\_30\_01 & LensKit & user\_knn.UserUser & explicit & nnbrs=30, min\_nbrs=30, min\_sim=0.1 \\ 
        UUEx\_60\_20\_0075 & LensKit & user\_knn.UserUser & explicit & nnbrs=60, min\_nbrs=20, min\_sim=0.075 \\ 
        UUEx\_120\_2\_001 & LensKit & user\_knn.UserUser & explicit & nnbrs=120, min\_nbrs=2, min\_sim=0.01 \\ 
        UUEx\_120\_30\_001 & LensKit & user\_knn.UserUser & explicit & nnbrs=120, min\_nbrs=30, min\_sim=0.01 \\ 
        FunkSVD & LensKit & funksvd.FunkSVD & explicit & damping=5, features=250, iterations=175,\newline lrate=0.001, reg=0.015 \\ 
        BiasedMF & LensKit & als.BiasedMF & explicit & features=250 \\ 
        ImplicitMF & LensKit & als.ImplicitMF & implicit & features=250 \\ 
        ADMMSLIM & RecBole & ADMMSLIM & implicit & alpha=1, lambda1=5, lambda2=1000,\newline epochs=1 \\ 
        BPR & RecBole & BPR & implicit & embedding\_size=2048, learning\_rate=0.0001, epochs=659 \\ 
        CDAE & RecBole & CDAE & implicit & reg\_weight\_1=0.01, reg\_weight\_2=0,\newline learning\_rate=0.05, epochs=176 \\ 
        EASE & RecBole & EASE & implicit & reg\_weight=500, epochs=1 \\ 
        MultiVAE & RecBole & MultiVAE & implicit & mlp\_hidden\_size=[300], dropout\_prob=0.3,\newline anneal\_cap=0.1, learning\_rate=0.01,\newline epochs=227 \\ 
        NeuMF & RecBole & NeuMF & implicit & ml\_hidden\_size=[256,128,256], dropout\_prob=0.2, learning\_rate=0.0005,\newline epochs=137 \\ 
        \bottomrule
    \end{tabular}
\end{table*}

To increase the variety of recommendations, many of our runs utilized
user and item knn methods with different parameters settings.  Some
parameter settings were too drastic and resulted in those
runs failing to produce recommendations for all participants.

Chamani~\cite{chamani24:thesis} provides full details on how we set
parameters for the various algorithms.

\subsection{Pooling}

Using each algorithm, we produced ranked lists of up to 1000
recommendations such that these movies were not in the participant's
full profile.  Thus, even though the implicit dataset does not contain
a participant's full profile, we filtered from the recommendations
produced by the algorithms the items already rated by the user's full
profile.

As mentioned earlier, relevance assessment involved two phases of
judging, and these phases differed by how deeply we pooled
recommendations.  Phase 1 had a minimum pool depth of 10, and phase 2
had a minimum pool depth of 50.  In addition, each phase had a minimum
number of movies to be rated.  For phase 1, if the number of movies to
be rated did not yet equal at least 135 movies, we continued further down
into the pool until we had at least 135 movies for the
participant.  For phase 2, the minimum number of movies to rate was
set at 600 in total (phase 1 + phase 2).    

\subsection{Consistency Checks}

In addition to the movies in the pool, for each participant we added
randomly selected movies from their ratings profile to the set of
movies to be judged.  We call these movies \textit{consistency
  checks}, for they allow us to compare the participants' ratings on
these items to their existing ratings in their profiles.  Prior
experience with crowdsource work and other recruited participants has
taught us that even though the remuneration amounts are low, they can
still be attractive to people who want to quickly enter fake
assessments rather than take the time to do careful work.

For phase 1, we added 10 randomly selected movies, and for phase 2, we
added 50 randomly selected movies.  If any of the consistency check
movies for phase 2 had already been selected as consistency checks for
phase 1, we utilized the phase 1 judgment.  As shown
by~\citet{chamani24:thesis}, we deemed all 51 participants to have
suitable consistency for inclusion in this extension to ML-32M.
Future work analyzing these judgments should be able to contribute to
the existing research on noise in ratings~\cite{amatriain09:noise,said12:magic,said18:magic}.

\subsection{Relevance Assessment}
\label{sec.assess}

We built a website application that allowed our participants to login
and judge their pools of movies.  Participants saw one movie at a
time. On submission of a movie assessment, we showed them the next
movie in their pool.  The display for a movie included a poster image,
the movie's title, a plot summary, release year, running time in minutes,
cast, director(s), genres, and language(s) of the movie.

Assessing a movie consisted of answering three questions, and for each
question, we provided a dropdown to allow the participant to select
their answer.  When we asked for ratings, we used the same rating scheme
as movielens.org, which included its text descriptions of what the
star ratings mean. The movielens.org rating scheme as we displayed it to
the participants:
\begin{itemize}
\item 0.5 stars (Awful)
\item 1 star (Awful)
\item 1.5 stars (Poor)
\item 2 stars (Poor)
\item 2.5 stars (OK)
\item 3 stars (OK)
\item 3.5 stars (Good)
\item 4 stars (Good)
\item 4.5 stars (Must Watch)
\item 5 stars (Must Watch)
\end{itemize}  
The three questions we asked:
\begin{enumerate}

\item ``How familiar are you with this movie?'' Answer choices:
  \begin{itemize}
    \item Unseen - Never heard of it
    \item Unseen - Familiar with movie
    \item Unseen - Very familiar (read reviews, seen trailers, etc.)
    \item Seen the movie - 0.5 stars (Awful)
    \item \ldots We repeat ``Seen the movie'' with each of the movielens.org star ratings as per above. \ldots
    \item Seen the movie - 5 stars (Must Watch)
  \end{itemize}

\item ``How interested are you in watching this movie via a streaming
  service?'' Answer choices: Not interested, Somewhat interested, 
  Interested, Very interested, Extremely interested.

\item ``For unseen movies only: If you were to watch this movie, what would
you predict your rating of it to be?'' Answer choices are the movielens.org star ratings as per above.

\end{enumerate}

In addition to the questions as shown in the web
application, the participants had read instructions and then answered
a quiz to ensure that they had read the instructions before beginning
assessment.  In our instructions to participants, we explained to them
that they should imagine they had been given a subscription to a
streaming service that had all of the movies available to watch for
free. We stressed that interest-to-watch held for seen and unseen
movies.  We explained that it is possible to want to watch already
seen movies again, while it is also possible that they may love a
movie but not be interested in watching it again.

Throughout the assessment process, the application also allowed
participants to view all of their assessments and edit each assessment
if needed.

After a participant had judged all of the movies in their pool, they
were then asked to select and rank their top 3 choices for watching
with their rank 1 choice being the movie they most wanted to watch
next, and rank 2 their next choice, and so forth.  Participants could
select movies from those they had rated as being ``Extremely
interested'' in watching, and if they had fewer than 3 such movies, we
also included the movies from the next preference level ``Very
interested'' and so forth.

\subsection{Qrels: New Evaluation Objectives}

From the collected relevance assessments, we are able to create many
different sets of relevance judgments (qrels in TREC parlance).  With
participants assessing movies on their preference for watching them, 
we have partial preference judgments where we have equivalence classes
of movies at different preference levels.  For example, there may be 7
movies that a participant judges as being ``Extremely interested'' in
watching, and 23 movies for which they are ``Very interested'' in
watching.  All 7 movies in the equivalence class ``Extremely
interested'' are preferred to all 23 movies in ``Very interested'',
but none of the 7 are preferred to each other.

As noted in Section~\ref{sec.assess}, participants also selected three
movies and ranked them as their rank 1, 2, and 3 movies for interest
in watching.  We thus have a preference ordering where there is one
movie at each of ranks 1-3 and then possibly multiple movies at each
of the remaining preference levels.  Movies that are assessed as
``Not interested'' are the equivalent of non-relevant.

A mistake that we made was to include in this final ranking step the
movies used for the consistency checks.  To our chagrin, we discovered
that sometimes participants most wanted to watch movies they had
already rated as part of their MovieLens profiles.  Because our
objectives are all cast as making recommendations for movies not
already in a user's rating profile, we exclude these movies from the
relevance judgments, and thus not all participants have a rank 1, 2,
and 3 most preferred movies.

Our primary qrels, interest.qrels, captures the participants' interest
in watching movies and has preference level values of 0 (Not
interested), 1 (Somewhat interested), 2 (Interested), 3 (Very
interested), 4 (Extremely interested), 5 (ranked 3), 6 (ranked 2), and
7 (ranked 1).  Like all qrels that we produced, we excluded the
consistency check movies.

With the familiarity information that we also collected from our
participants, we are also able to generate a host of other preference
qrels.  These are described in Table~\ref{tab.interest.qrels}.  These
preference qrels, are designed to be used with the
\textit{compatibility} measure~\cite{clarke21:compat_tois}, which
reports a measure of how well a ranked list matches an ideal ordering
of items given the preference levels recorded in the qrels. For our
experiments, we use the version of
compatibility\footnote{\url{https://github.com/trec-health-misinfo/Compatibility}}
used with the TREC Health Misinformation track~\cite{healthmisinfo20},
which differs from the official
release\footnote{\url{https://github.com/claclark/Compatibility}} in
that it outputs in a trec\_eval style and only reports on topics in
the qrels. The \textit{compatibility} measure ignores the preferences
given a value $\le 0$ in a qrels file, and thus while we included
``Not interested'' movies in interest.qrels, they are treated the same
as unjudged movies.  The bad-not-interested.qrels and the
unheard-high-rating.qrels only contains judgments from 48 and 43
participants, respectively.  These two qrels files contain fewer
participants in them because some participants had no judgments that
met the criteria of the file, for example, 8 participants did not rate
any unheard-of movies $\ge 4$. 

In addition to their interest in watching movies, participants also
provided a predicted MovieLens-scale rating for unseen movies and an
actual rating for already seen movies that were recommended to them
but not in their submitted profiles.  We have also created qrels from
these ratings, but the primary purpose of these qrels is for
comparison to the interest-based preference judgments.  Several
participants mentioned that it was difficult to make predicted
ratings, and as such, the interest-based qrels are preferred.  The
ratings-based qrels are described in Table~\ref{tab.rating.qrels}.

\begin{table*}
\caption{Preference-based relevance judgments (qrels) designed to be
  used with the \textit{compatibility} measure.}
\label{tab.interest.qrels}
\begin{tabular}{lp{4.1in}}
\toprule
Qrels & Objective \\ \midrule
interest.qrels & Primary objective to maximize. Use with \textit{compatibility} measure. Preference levels for interest in watching movies. A higher value is preferred to a lower value. \\
seen-interest.qrels & Same as interest.qrels, but only for movies already seen by the participant. \\
unseen-interest.qrels & Same as interest.qrels, but only for movies unseen by the participant. \\ 
not-very-familiar-interest.qrels & Same as unseen-interest.qrels, but only for movies with familiarity less than ``very familiar''. \\
unheard-interest.qrels & Same as interest.qrels, but only for movies the participant reports zero familiarity. \\
interest-prefer-less-familiar.qrels & This maintains the preference levels of interest.qrels, but within each preference level, less familiar movies are preferred to more familiar movies. \\
bad-not-interested.qrels & All ``not interested'' movies with a
predicted MovieLens rating $\le 2.0$ (bad or awful).  A recommendation
algorithm wants to \textit{minimize} compatibility with these qrels. \\
\bottomrule
\end{tabular}
\end{table*}

\begin{table*}
\caption{Ratings-based relevance judgments (qrels) to be
  used for prediction accuracy or with a measure such as nDCG.}
\label{tab.rating.qrels}
\begin{tabular}{lp{4.45in}}
\toprule
Qrels & Objective \\ \midrule
rating.qrels & The participant's predicted MovieLens rating for unseen movies, and their actual rating for seen movies. To be used for prediction accuracy measures. Not to be used for measuring ranking effectiveness with nDCG. \\
seen-rating.qrels & Same as rating.qrels, but only for movies already seen by the   participant. Not with nDCG. \\
unseen-rating.qrels & Same as rating.qrels, but only for movies unseen by the participant. Not with nDCG. \\
unheard-rating.qrels & Same as rating.qrels, but only for movies the participant reports zero familiarity. Not with nDCG. \\
high-rating.qrels & Same as rating.qrels, but only retains ratings 4.0 and greater. It and derivatives may be used with nDCG, but in general, the corresponding preference-based qrels in Table~\ref{tab.interest.qrels} should be used for ranking. \\
seen-high-rating.qrels & Same as high-rating.qrels, but only for seen movies. Okay with nDCG. \\
unseen-high-rating.qrels & Same as high-rating.qrels, but only for unseen movies. Okay with nDCG. \\
unheard-high-rating.qrels & Same as high-rating.qrels, but only for movies the participant reports zero familiarity. Okay with nDCG. \\
\bottomrule
\end{tabular}
\end{table*}

\section{Comparisons}
\label{sec.comparison}

In this section, we investigate how our pooling approach and new
objective functions compare to the traditional train/test evaluation
approach for recommendation systems, and in particular, for evaluation
using MovieLens-32M (ML-32M).

\subsection{Traditional Train/Test}
\label{sec.train.test}

To be able to compare evaluation using our pooling-created ML-32M
extension to a traditional train/test split, we created a 80\% train
and 20\% test split of 10K existing user profiles and our 51
participants.  We randomly selected the 10K existing MovieLens users
from the non-participant profiles with at least 20 ratings in 
our implicit-ratings dataset.  Our train/test split is a random
stratified sample.  We first take a profile and divide it into the
ratings $\ge 4.0$ and those $< 4.0$.  For the ``high'' ratings
$\ge 4.0$, we then do a random 80/20 train/test split, and likewise do the
same for the ``low'' ratings $< 4.0$.  We then recombine the low and
high train ratings, and the low and high test ratings.  We use the
stratified sample to be sure that we can create qrels for high-ratings
$\ge 4.0$ and so that our qrels for a full ratings profile contains
equal samples of loved and not-loved movies.  There are many other
ways to perform a train/test traditional evaluation, but the simple
random split remains popular~\cite{sun23:fresh}.

Using the test ratings, we produced qrels separately for the random 10K
MovieLens users, our 51 participants, and for both groups, we produced
qrels using all ratings values (all-ratings) and only ``high'' ratings
$\ge 4.0$ (high-ratings).
 
\subsection{Runs and Evaluation Measures}

With the train/test splits of our explicit and implicit datasets
(Section~\ref{sec.data}), we use the same 22 algorithms as described in
Table~\ref{tab.algs} to produce recommendations for our random-10K
users and our 51 participants.  In all cases, our recommendations
exclude all ratings from a user's or participant's explicit training
data, and thus recommendations using the implicit dataset also exclude
known training data in the explicit train set.

To evaluate our runs, we modified LensKit and RecBole to produce
recommendations in TREC results format, which then allowed us to use
trec\_eval\footnote{\url{http://github.com/usnistgov/trec_eval}} and
compatibility.py to compute nDCG@100 and \textit{compatibility}
(p=0.98) scores, respectively, with our TREC format qrels.  We use
nDCG@100 with ratings-based qrels, and \textit{compatibility} (p=0.98) with
preference-based qrels.

\subsection{Observations}

With a new dataset such as ours, there is more than can be
investigated and shared in one paper.  We focus our investigation on
the differences in evaluation caused by selection of user profiles
(participants vs.\ random ML-32M users), differences between using the
interested-in-watching preferences and using ratings, and the
differences caused by using our pooling-based evaluation and the
traditional train/test split. 

\subsubsection{Participants vs.\ 10K Random ML-32M Users}

As noted in Section~\ref{sec.participants}, our participants appear to
be movie enthusiasts who have created some of the larger profiles in
ML-32M while the random ML-32M profile is much smaller.  We can get a
sense of the impact on evaluation caused by using our 51 participants
rather than random ML-32M users by comparing how each set of profiles
affects the evaluation of our 22 different recommendation runs on the
train/test setup.  For the train/test setup, we have qrels for both
sets of profiles, and for each we have test all-ratings, and a set of
test high-ratings ($\ge 4.0$).

\begin{figure}
  \centering
  \includegraphics[width=4in]{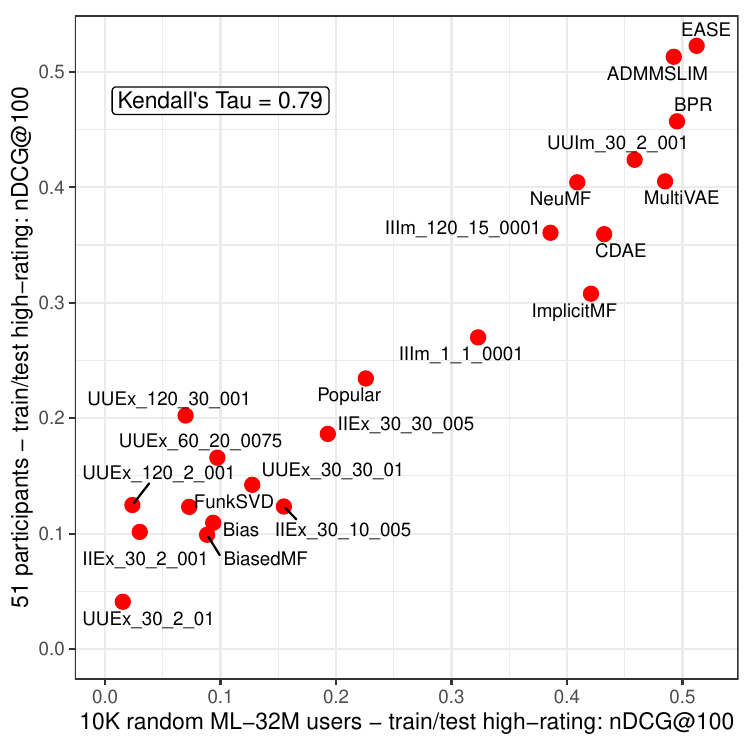}  
  \caption{Traditional train/test evaluation with 51 participants
    vs.\ 10K ML-32M users for ratings $\ge 4.0$.  This figure shows
    the effect of using the 51 study participants rather than a random
    sample of ML-32M users.  Both axes represent the same train/test
    evaluation approach but each axis has a different set of user
    profiles.  We can see that for the top half of runs, there are
    small changes in the ranking of the algorithms between our 51
    participants and the random 10K ML-32M users.  The majority of the
    rank changes are occurring with the lower performing algorithms.
    With a traditional train/test evaluation approach the Popular
    algorithm is ranked in the middle of the 22 runs.
  }
  \label{fig.part.v.rand}
  \Description{A scatter plot showing the evaluation of the 22 runs
    used for building the judging pools.  For both axes, the
    evaluation is with a traditional train/test approach.  The
    vertical axis shows the nDCG@100 scores when evaluation is via the
    51 study participants and using high ratings, i.e.\ where ratings
    are greater than or equal to 4.0.  The horizontal axis shows the
    same but for the ten thousand randomly selected ML-32M users. We
    can see that for the top half of runs, there are small changes in
    the ranking of the algorithms between our 51 participants and the
    random 10K ML-32M users.  The majority of the rank changes are
    occurring with the lower performing algorithms.  The Kendall's tau
    is 0.79. The top-performing run is EASE and in the middle of the
    runs is Popular.}
\end{figure}

We computed Kendall's $\tau$ to measure the correlation between
ranking runs with our participants' train/test profiles vs.\ the
random 10K ML-32M users we selected as per
Section~\ref{sec.train.test}.  The correlation on high-ratings was the
highest at 0.79, and the correlation on all-ratings was slightly lower
at 0.77.  Figure~\ref{fig.part.v.rand} shows the 51 participants
vs.\ the 10K ML-32M users using high-ratings, and while not shown, the
plot using all-ratings is very similar.

Figure~\ref{fig.part.v.rand} shows that there is a difference between
evaluating with our participants rather than random ML-32M users, and
the correlations of 0.77-0.79 are below the oft-cited
\citet{voorhees98:variations} threshold of 0.9 for declaring that two
methods are ranking runs similarly, but this is not a bad difference.
First, the majority of the rank changes for the evaluated runs appear
to be occurring in the lower half of the runs, while the top
performing runs see only small changes in rank.  Secondly, and perhaps
most importantly, it can be argued that our participants represent a
more suitable user scenario to optimize for than the random ML-32M
user profile.  Our participants appear to be active users of
movielens.org, and are probably the movielens.org users who most make
use of recommendations to find movies.  A significant portion of
random ML-32M users visited the site, rated at least 20 movies, and
then did not become regular users.  While producing good
recommendations for new users is a valid research problem, it is a
different problem than recommending movies to regular users.

\subsubsection{Interested-in-Watching Preferences vs.\ Ratings}

\begin{figure}
  \centering
  \includegraphics[width=4in]{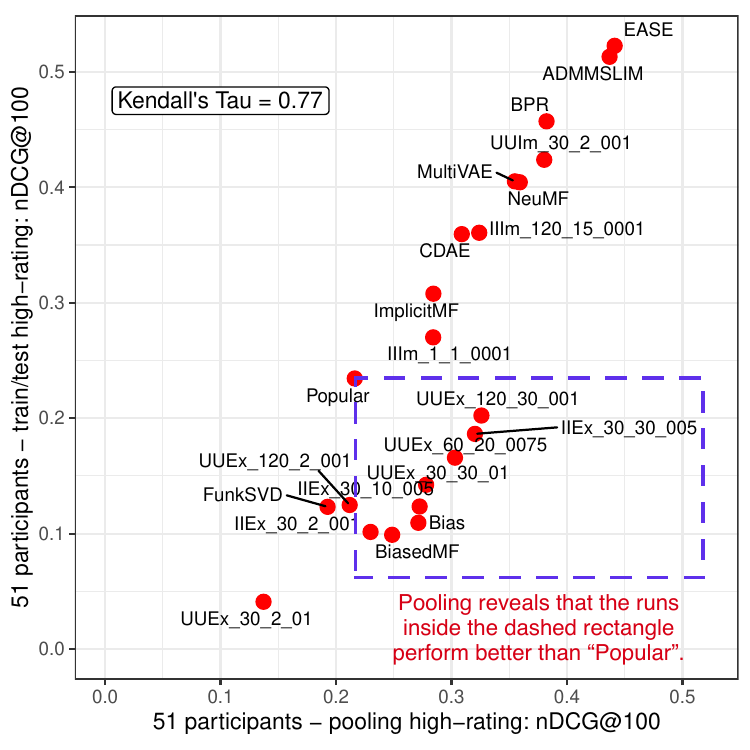}  
  \caption{Traditional train/test evaluation vs.\ pooling based
    evaluation.
    Each axis shows an evaluation with the same 51 participants.
    The vertical axis shows nDCG@100 scores for the high ($\ge 4.0$)
    test ratings in a train/test split evaluation.
    The horizontal axis shows the pooling-based evaluation with
    nDCG@100 scores for high-rating.qrels ($\ge 4.0$).
    Of note, the Popular algorithm has dropped from rank
    11 with train/test evaluation to rank 19 with pooling-based
    evaluation.}
  \label{fig.cran.v.trad}
  \Description{A scatter plot showing the evaluation of the 22 runs
    used for building the judging pools that compares pooling based
    evaluation vs.\ traditional train/test evaluation.  Each axis
    shows an evaluation with the same 51 participants.
    The vertical axis shows nDCG@100 scores for the high ($\ge 4.0$) test ratings in a train/test split
    evaluation.  The horitzontal axis shows the pooling-based evaluation with nDCG@100 scores for
    high-rating.qrels ($\ge 4.0$).  Of note, the Popular algorithm has dropped from rank
    11 with train/test evaluation to rank 19 with pooling-based
    evaluation. The Kendall's tau is 0.77.}
\end{figure}

We next turn our attention to our pooling-based evaluation and ask
about the difference between using the interest-in-watching
preferences, e.g.\ interest.qrels in Table~\ref{tab.interest.qrels},
and the ratings-based qrels, e.g. rating.qrels in
Table~\ref{tab.rating.qrels}.  When we compared all pairs of rank
correlations between rankings of recommendation runs using the two
sets of qrels (Table~\ref{tab.interest.qrels}
vs.\ Table~\ref{tab.rating.qrels}), we found that interest.qrels and
high-rating.qrels had a correlation of 0.96, and thus rank runs
similarly.  It appears that
participants' preferences for what they want to watch correlates well
with the movies they predict they will rate $\ge 4.0$ once viewed or
those they have already seen and want to see again.

We also found that unseen-interest.qrels and high-rating.qrels had a
high correlation of 0.91. We found seen-interest.qrels to have a
correlation of 0.88 with both seen-rating.qrels and
seen-high-rating.qrels. Likewise, unheard-interest.qrels had a
correlation of 0.87 with unheard-high-rating.qrels, and 0.86 with
unheard-rating.qrels.

The highest correlation for rating.qrels and an interest-based qrels
was with interest.qrels at only a correlation of 0.67.  Thus we see
that the inclusion of movies rated less than 4.0 significantly changes
the evaluation of runs compared to using preferences for watching.
While preferences for interest-in-watching reward runs for getting preferred
movies near the top of the recommendations, using all ratings for
evaluation with nDCG will reward runs even for lower rated movies, for
which there may be little or no interest in watching.  We recommend
against using all rating values with nDCG for measuring the
effectiveness of top-n recommendations.  The rating.qrels should only
be used for rating prediction, e.g.\ mean absolute error (MAE).  This
finding is in line with \citet{breese98:empirical} who set to zero all
ratings less than or equal to the ``neutral'' rating when measuring
expected utility of a ranked list. 

\subsubsection{Pooling-based (Cranfield) vs.\ Train/Test Split
  Profiles}

As noted above, our pooling created interest-in-watching
(interest.qrels) and the participants' predicted ratings (and actual
ratings for seen movies) $\ge 4.0$ (high-rating.qrels) effectively
rank recommendation systems the same (Kendall's $\tau$ = 0.96).  Thus, we
can compare our Cranfield, pooling-based evaluation approach to a
machine learning styled approach with its train/test split of ML-32M
profiles by comparing our 51 participants' pooled high-rating with the 51
participants' train/test high-rating.  We earlier compared
participants to the 10K random ML-32M users in Figure~\ref{fig.part.v.rand}.

Figure~\ref{fig.cran.v.trad} shows a traditional train/test split
evaluation vs.\ our pooling-based evaluation.  The effect of how the
test collection is built is isolated, for 1) the effect of our
participants being different from random ML-32M users is removed
because both evaluations are with our 51 participants, and 2) the
effect of preferences vs.\ ratings is removed because both use ratings
$\ge 4.0$ and nDCG@100.  The most significant change in evaluation is
that the Popular algorithm has dropped from being in the middle of the
pack (rank 11 of 22 runs) to near the bottom (rank 19).  The other
runs have an average absolute change in rank of 2.2, and Popular has
the most extreme change of 8.

As Figure~\ref{fig.cran.v.trad} shows, 
the top four runs do not change their rank order, and as noted, the
majority of changes in rank are minor, and this shows that the
traditional train/test split is not broken in a manner that prevents
it from generally identifying the better recommender system.
Nevertheless, with the change in Popular's rank, we see good evidence
that using existing ratings as relevance judgments has a popularity
bias and that our extension to ML-32M offers a solution to this
problem.

\begin{figure}
  \centering
  \includegraphics[width=4in]{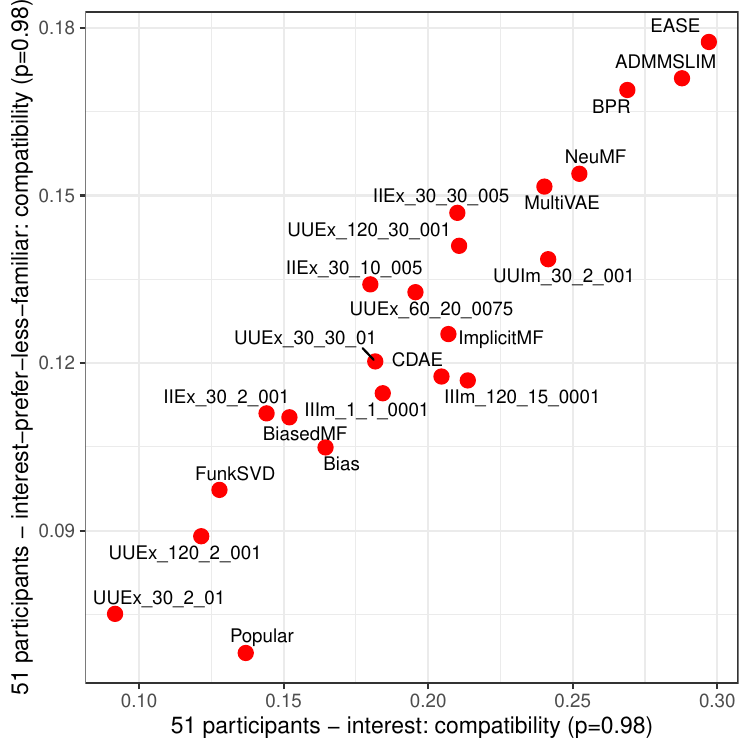}  
  \caption{Pooling-based evaluation with our 51 participants on both
    axes using preferences and the \textit{compatibility} (p=0.98)
    measure. The vertical axis shows the performance of the runs when
    measured with the interest-prefer-less-familiar.qrels and the
    horizontal shows the regular interest.qrels.  Of note, the Popular
    run becomes the worst of all runs when we score by interest in
    watching, and within preference levels, we prefer less familiar
    movies.}
  \label{fig.prefer.less.fam}
  \Description{A scatter plot showing the evaluation of the 22 runs
    used for building the judging pools that compares pooling based
    evaluation with two different sets of interest-based qrels and the
    \textit{compatibility} (p=0.98) measure.  Each axis shows an
    evaluation with the same 51 participants.  The vertical axis shows
    the performance of the runs when measured with the
    interest-prefer-less-familiar.qrels and the horizontal shows the
    regular interest.qrels.  Of note, the Popular run becomes the
    worst of all runs when we score by interest in watching, and
    within preference levels, we prefer less familiar movies.}
\end{figure}

We collected interest-in-watching preferences as well as a
participant's familiarity with a movie.  An often stated goal of
recommender systems is to help people find unfamiliar items that they
will enjoy.  We combined interest-in-watching with preference for less
familiar items by ordering the preference of items within each
interest-in-watching preference level from least to most familiar.
Figure~\ref{fig.prefer.less.fam} shows the effect of preferring less
familiar movies while still maintaining the overall interest-in-watching
preferences.  With this objective, the Popular run is now the lowest
performing run.  Interestingly, we also see more separation between
the top performing runs, the middle of pack is shuffled, and the worst
performing runs become clear.  While we do not know if users would
prefer less familiar items, this objective may be useful for its
ability to apparently remove popularity bias.

\section{Recommended Usage of ML-32M-Extension}
\label{sec.recs.for.use}

We recommend the usage of the interest-in-watching preferences
(interest.qrels) combined with the \textit{compatibility} measure with
the persistence, p, set appropriately for the evaluation scenario.
Setting p=0.95 (the default) is appropriate for evaluations interested
in evaluation that emphasizes the top 20 results, and p=0.98 for
evaluation depths around 50-100.  The preferences should not be
treated as relevance grades, and are not measures of gain.

To investigate a recommender's ability to find movies users are
interested in watching while promoting less familiar fare, the
interest-prefer-less-familiar.qrels are suitable, but these have not
yet been validated with user studies as matching what users prefer.
The same goes for using the preferences in seen-interest.qrels,
unseen-interest.qrels, etc.  The qrels other than interest.qrels are
useful to examine the ranking behavior of an algorithm, but should not
be the primary objective, for our participants did not express their
preferences in this fashion.  Our participants gave us their
preferences for interest-in-watching movies, and in that preference
ordering, captured all features such as seen and unseen.

For example, if we look at unheard-interest.qrels, the best run is
UUEx\_30\_2\_01 as measured by \textit{compatibility} (p=0.98), but
this run is the worst run when we measure interest-in-watching using
interest.qrels.  We would not want to optimize solely for
unheard-interest, for it would promote runs like UUEx\_30\_2\_01 that
perform horribly on our primary objective.  

The bad-not-interested.qrels can be used as an objective to minimize,
i.e.\ an algorithm does not want to score highly with these qrels, for
a high score means that bad recommendations are present in the ranked
list. 

The rating based qrels in Table~\ref{tab.rating.qrels}
should be used for testing an algorithm's ability to predict actual
ratings, and not for assessing top-n ranking ability.  For ranking,
the interest.qrels with \textit{compatibility} should be preferred.  

\section{Concluding Discussion}

Using established methods for building an IR test
collection~\cite{sanderson10:testcol}, we extended the ML-32M dataset
with 51 user profiles and preference judgments for their
interest-in-watching movies.  This was possible because top-n
recommendation is information retrieval.  For our extension to ML-32M,
each profile represents the context of a search topic where the
information need is ``recommend me unrated movies that I would be
interested in watching.''  By creating several different sets of
relevance judgments, we are able to represent other information needs
such as ``recommend me unrated movies that I have not already seen and
would be interested in watching.''

We believe that when feasible, offline test collections should have
the user with the information need be the person to assess the
relevance of the retrieved items.  Furthermore, capturing preferences
for items is to be preferred to relevance grades or
ratings~\cite{clarke21:compat_tois}.

By following a traditional IR test collection methodology, we can
argue and see that this methodology reduces popularity bias in offline
recommender systems evaluation.  The argument is simple: our study
participants assessed their recommendations for their interest in
watching the movies.  Their preference for one movie over another
captures all possible factors that may have gone into their decision.
If a participant prefers popular movies, then they should be
recommended popular movies and vice versa.  We should not as designers
of recommendation systems declare whether or not popular movies are
good recommendations.  We can also see that this methodology reduces
popularity bias in evaluation by the low performance of the Popular
algorithm in Figure~\ref{fig.cran.v.trad}.

Future work calls for more recommender systems test collections
to be constructed using IR test collection construction techniques, and in
particular, for these collections to be built as a group effort as
part of TREC, CLEF, NTCIR, FIRE, etc., for limitations of our work
include our selection of algorithms and our transformation of ML-32M
prior to producing recommendation pools.  A group effort would start
with the full dataset and user profiles being made available to all
participants, and thus it would be possible for any of the items in
the collection to be recommended.  Unfortunately, because we applied
k-core filtering, the movies we eliminated from ML-32M had no chance
of being in the pools.  Similarly, while we used a diverse set of
algorithms to produce recommendations, our effort pales with the
diversity that can be obtained from having different research groups
contribute their best runs.  As such, our work should be viewed as a
pilot for larger, better efforts at test collection construction.

\begin{acks}

Special thanks to Joseph Konstan and Daniel Kluver for helping us to
recruit movielens.org users and for their creation of the
ML-32M dataset.  We are very grateful for the advice and assistance of
Michael Ekstrand.  Thanks to the creators of the RecBole framework for
both its creation and their help with our questions.  Thanks to Abdul
Manaam for his assistance with data cleaning.  Thanks to Charles
Clarke for his feedback.  This research was supported in part by the
Natural Sciences and Engineering Research Council of Canada
(RGPIN-2020-04665, RGPAS-2020-00080).

\end{acks}

\bibliographystyle{ACM-Reference-Format}
\bibliography{recsys-eval}

\end{document}